\documentclass{elsart}

\usepackage{graphicx}

\usepackage{amssymb}

\begin{document}

\begin{frontmatter}

\title{Competition between local potentials and attractive particle-particle 
interactions in superlattices}

\author[univasf]{N. A. Lima}
\author[unesp]{A. L. Malvezzi}
\author[ifsc]{K. Capelle}
\ead{capelle@ifsc.usp.br}
\ead[url]{http://www.ifsc.usp.br/\~\,dft}

\address[univasf]{Colegiado de Engenharia de Produ\c{c}\~ao,         
Funda\c{c}\~ao Universidade Federal do Vale do S\~ao Francisco, 
Caixa Postal 252, 56306-410 Petrolina, PE, Brazil}

\address[unesp]{Departamento de F\'{\i}sica,
Faculdade de Ci\^encias, 
Universidade Estadual Paulista,
Caixa Postal 473, 17015-970 Bauru, SP, Brazil}

\address[ifsc]{Departamento de F\'{\i}sica e Inform\'atica,
Instituto de F\'{\i}sica de S\~ao Carlos,
Universidade de S\~ao Paulo,
Caixa Postal 369, 13560-970 S\~ao Carlos, SP, Brazil}

\begin{abstract}
Naturally occuring or man-made systems displaying periodic spatial modulations
of their properties on a nanoscale constitute superlattices. Such
modulated structures are important both as prototypes of simple 
nanotechnological devices and as particular examples of emerging spatial
inhomogeneity in interacting many-electron systems. Here we investigate the
effect different types of modulation of the system parameters have on the 
ground-state energy and the charge-density distribution of the system. The 
superlattices are described by the inhomogeneous attractive Hubbard model, 
and the calculations are performed by
density-functional and density-matrix renormalization group techniques.
We find that modulations in local electric potentials are much more effective 
in shaping the system's properties than modulations in the attractive on-site 
interaction. This is the same conclusions we previously (Phys. Rev. B 
{\bf 71}, 125130) obtained for repulsive interactions, suggesting that 
it is not an artifact of a specific state, but a general property of 
modulated structures.
\end{abstract}

\end{frontmatter}

\newcommand{\be}{\begin{equation}}
\newcommand{\ee}{\end{equation}}
\newcommand{\bea}{\begin{eqnarray}}
\newcommand{\eea}{\end{eqnarray}}
\newcommand{\bi}{\bibitem}

\renewcommand{\r}{({\bf r})}
\newcommand{\rp}{({\bf r'})}

\newcommand{\ua}{\uparrow}
\newcommand{\da}{\downarrow}
\newcommand{\la}{\langle}
\newcommand{\ra}{\rangle}
\newcommand{\dg}{\dagger}

\section{Introduction}
\label{intro}

Superlattices are characterized by a periodic modulation of system properties
on a length scale that is larger than the lattice constant of the material
they are made of. Superlattices are of great technological interest because 
the modulation can be used to design the system's optical, electrical, magnetic
and transport properties \cite{superl1,superl2,superl3,superl4,superl5}. 
Since the modulation typically takes place on a 
length scale of nanometers, superlattices also provide a particular realization
of systems with nanoscale spatial inhomogeneity, which occurs spontaneously in 
many strongly correlated systems \cite{inhom1,inhom2,inhom3}.

A simple, but nontrivial, model of a superlattice is the inhomogeneous 
one-dimensional Hubbard model (1DHM) for spin $1/2$ fermions on
a lattice, with Hamiltonian
\be
\hat{H}_{inhom}=
-t \sum_{i,\sigma} (c_{i\sigma}^\dagger c_{i+1,\sigma} + H.c.)
+\sum_i U_i c_{i\ua}^\dagger c_{i\ua}c_{i\da}^\dagger c_{i\da}
+\sum_{i\sigma} v_i c_{i\sigma}^\dagger c_{i\sigma},
\label{1dhm}
\ee
where either the on-site interaction $U_i$, or the on-site potential $v_i$
or both are periodically modulated. Both for designing artificial superlattices 
and for the study of spontaneously emerging periodic structures, it is 
important to know which of these two types of modulations is more effective 
in shaping the system properties. 

In Ref.~\cite{superlatticeprb} we investigated this question for repulsive 
interactions, and found that the properties of the superlattice are much more 
sensitive to local changes in the electric potential than to local changes 
in the strength of the interparticle interaction. This is a
surprising result, as intuitively one would expect that in a strongly
correlated system, such as described by the 1DHM, interaction effects
should dominate. Indeed, in most applications of the Hubbard model (not
only to superlattices, but also, e.g., to HTSC cuprates) the on-site potential 
is not even considered, or simply assumed to be spatially constant and
absorbed in the chemical potential.
In the present paper we extend this investigation to attractive interactions,
i.e., $U<0$. The negative $U$ Hubbard model has been used to describe 
superconductivity of strongly correlated electrons, but the same model also 
describes attractive interactions between cold atoms in optical traps
\cite{coldatoms}.

\section{Comparison of modulated structures}
\label{criteria}

Many different modulation patterns exist, and many different ways to compare 
them are conceivable. What we require for our present purposes are
quantitative criteria measuring the relative impact of modulations in
the on-site potential, as compared to modulations in the on-site interaction.

In Ref. \cite{superlatticeprb} we introduced two such criteria, which are 
both based on separate calculations for a 1DHM with no modulations (a 
spatially uniform system) and for a 1DHM in which $U_i$ and $v_i$ are 
modulated with same periodicity but different amplitudes. In the present paper,
dedicated to attractive interactions, we choose a periodically repeated 
pattern of $L_U$ sites with interaction $U_i<0$ and potential $v_i$, 
followed by $L_0$ sites with $U_i=v_i=0$. The modulations of $U_i$ and of
$v_i$ have distinct effects on the system properties. The ratio of the 
modulation amplitudes at which the two effects cancel each other, returning
effectively to the properties of the uniform system, provides a quantitative 
measure of their relative impact on the system properties.

Two important ground-state properties of a modulated structure are its total
energy and its density profile. These quantities form the basis for the two 
criteria we employ. In the {\em energy criterium}, we vary the on-site 
potential until
the doubly modulated lattice has the same ground-state energy found in the
uniform system. The {\em density criterium} consists in finding that potential
that produces the same density profile obtained in the open uniform system.
Exact cancellations are harder to obtain for the density than for the energy, 
but cancellations to within the error bars of our numerical methods 
are all that is required for assessing the relative impact of modulating 
$v_i$ and $U_i$.

We stress that we do not expect such cancellations to occur
frequently in nature. We simply use them here as a convenient way to
quantify the relative impact of both types of modulation.
Essentially, we are simulating numerically the sensitivity
of the energy and the density to changes in the interaction and the
potential, i.e., sampling certain response functions.

\section{Computational methods}
\label{methods}

We calculate the energies and densities by means of two independently 
developed and implemented numerical techniques: the density-matrix 
renormalization group (DMRG) \cite{dmrg1,dmrg2} and density-functional
theory (DFT) within the Bethe-Ansatz local-density approximation (BA-LDA)
\cite{balda1,balda2,balda3,balda4}. Both methods provide energies and
density profiles of the Hamiltonian (\ref{1dhm}), but are complementary
in scope. DMRG provides high-precision results, but at considerable
computational expense, in particular for systems with reduced spatial
symmetries or periodic boundary conditions. BA-LDA easily handles systems of
almost arbitrary size and its performance does not significantly depend on
presence or absence of spatial symmetries or choice of boundary conditions.
Its intrinsic accuracy is, however, limited to a few percent (with 
energies typically being slightly more accurate than densities)
and it is not easy to consistently improve on these values.

Within density-functional theory, the ground-state energy and density
distribution are obtained by self-consistent solution of a set of 
single-particle equations, the Kohn-Sham equations, in the presence of 
effective potentials constructed such as to recover the density distribution 
of the interacting system. The application of DFT to the Hubbard model was 
initiated in Refs. \cite{gs}. Recent progress is due to the development of 
a simple and accurate Bethe-Ansatz based parametrization for the model's 
correlation energy in the uniform
limit, which can be used to construct a local-density approximation (LDA)
\cite{balda1,balda2,balda3}. This parametrization was further developed
into a fully numerical scheme \cite{balda4}, which has been applied both to
repulsive \cite{balda4} and attractive \cite{coldatoms} interactions, in the
context of ultracold fermionic atoms confined in optical traps. In the
same context, a simple analytical parametrization of the LDA for attractive
interaction was also proposed \cite{tlda}, but applied only in conjunction
with a similar LDA for the kinetic energy, leading to a Thomas-Fermi-like
scheme \cite{tlda}.

\begin{table}
\caption{\label{table1}
Ground-state energy of superlattices with open boundary conditions,
$L$ sites and $N$ electrons, subjected to spatially modulated on-site
interaction $U_i$ and spatially constant on-site potential $v_i=0$, obtained
with DMRG and with DFT/BA-LDA. The modulation pattern alternates $L_U$
sites of interaction $U<0$ with $L_0$ sites with interaction $U=0$,
simulating a superlattice composed of alternating superconducting and
normal metals. The last column is the deviation of the BA-LDA energies
from the DMRG ones (in percent).}
\begin{center}
\begin{tabular}{c|c|c|c|c|c|c|c}
$L$ & $N$ & U &  $L_U$ & $L_0$ & $E_0^{\rm DMRG}/t$ & $E_0^{\rm BA-LDA}/t$ &
$\Delta \% $ \\
\hline
60  & 30 & -3 & 60 &  0 & -72.989 & -74.118 & 1.55 \\
60  & 30 & -3 &  3 &  2 & -69.473 & -70.379 & 1.30 \\
100 & 40 & -3 &  1 &  1 & -90.059 & -90.740 & 0.76 \\
100 & 40 & -6 &  1 &  1 &-134.25  &-138.08  & 2.85 \\
300 &150 & -3 & 10 & 10 &-337.92  &-340.88  & 0.88 \\
300 &150 & -6 & 10 & 10 &-515.18  &-532.79  & 3.42 \\
300 &300 & -3 & 10 & 10 &-556.07  &-560.41  & 0.78 \\
\end{tabular}
\end{center}
\end{table}

\begin{figure}
\begin{center}
\includegraphics[height=90mm,width=70mm,angle=90]{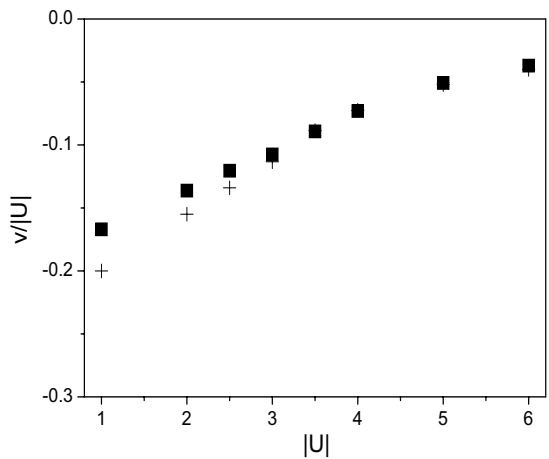}
\end{center}
\caption{\label{fig1} Amplitude $v$ of the modulation of the on-site potential
that produces the same ground-state energy in a superlattice whose interaction
is already modulated with amplitude $U$, as obtained in an unmodulated
system with interaction $U$ on all sites and $v_{unif}\equiv 0$.
Full squares: DMRG data. Plus: DFT/BA-LDA data. System parameters: $L=60$
sites, $N=30$ fermions, $L_U=4$ interacting sites, alternating with $L_0=6$
noninteracting sites.}
\end{figure}

Here we employ the parametrization for the correlation energy used in
Ref. \cite{tlda}, in conjunction with the Kohn-Sham approach of
Refs. \cite{balda1,balda2,balda3}, to obtain many-body energies and densities.
In these calculations a local-density approximation is made only for the 
correlation energy. Previous work \cite{coldatoms}, as well as our present 
results, indicate that the LDA
(in either its fully numerical or parametrized forms) for attractive
interactions continues to produce energies within a few percent of more
precise DMRG data. This is illustrated in Table \ref{table1}. DMRG densities
are less well reproduced by LDA densities at attractive than at
repulsive interactions, but Figs.~\ref{fig1} and \ref{fig2} show that
the differences are far too small to invalidate our conclusions regarding
the effect of different modulation patterns.

In view of the complementary aspects of DFT and BA-LDA , we have, both in
Ref. \cite{superlatticeprb} and in the present paper, used BA-LDA to
efficiently explore large regions of parameter space.
In separate calculations we have then used DMRG to fine-tune parameters
in the vicinity of the values provided by BA-LDA, and to confirm results with
more significant digits. The DMRG calculations were performed using the 
finite system algorithm under open boundary conditions \cite{dmrg1} with
truncation errors kept of the order of $10^{-6}$ or smaller.

\section{Attractive Hubbard model: results}
\label{attractive}

Figure \ref{fig1} displays representative results from the energy criterium. 
In these calculations we have chosen a fixed modulation of the interaction, 
alternating 'layers' of $L_U=4$ sites with $U_i=U$ and 'layers' of $L_0=6$ 
sites with $U_i=0$, and changed the amplitude of the modulation of $v_i$ 
until the doubly modulated system has the same energy as the unmodulated 
system (with interaction $U$ on all
sites and $v_{unif}\equiv 0$). If a different constant value of $v_{unif}$ 
is chosen in the uniform system, the compensating values of $v_i$ in the doubly
modulated system are simply shifted up by that value of $v_{unif}$, without 
changing our conclusions.

Figure \ref{fig1} shows that for physically realistic values of $U$
the effect of a modulation of $U$ on the ground-state energy is reproduced
by a modulation of $v$ that is between $10^{-2}$ and $2 \times 10^{-1}$
times smaller. Clearly, the ground-state energy depends much more 
sensitively on modulations of the on-site potential than on modulations
of the on-site interaction. Qualitatively, this is the same conclusion
obtained previously for repulsive interactions \cite{superlatticeprb}.
Quantitatively, the effect of attractive interactions is even weaker than
that of repulsive interactions, as can be seen by comparing the present 
Fig. \ref{fig1} with the upper panel of Fig. 4 of Ref. \cite{superlatticeprb}:
an even weaker modulation of the potential is sufficient to change the
energy by the same margin as a modulation of the interaction.

\begin{figure}
\begin{center}
\includegraphics[height=90mm,width=70mm,angle=90]{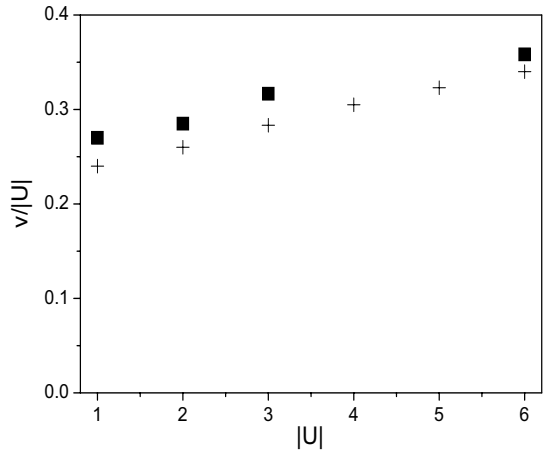}
\end{center}
\caption{\label{fig2} Amplitude $v$ of the modulation of the on-site potential
that produces the same ground-state density distribution in a superlattice 
whose interaction is already modulated with amplitude $U$, as obtained in 
an unmodulated system with interaction $U$ on all sites and 
$v_{unif}\equiv 0$.
Full squares: DMRG data. Plus: DFT/BA-LDA data. System parameters: $L=160$
sites, $N=80$ fermions, $L_U=6$ interacting sites, alternating with $L_0=10$
noninteracting sites.}
\end{figure}

In Fig. \ref{fig2} we turn to the density criterium, i.e., investigate for
what amplitude of modulation of an repulsive on-site potential $v_i$ it 
cancels the effect of a modulation of the attractive on-site interaction
$U_i$ on the density profile. The density is more sensitive to small 
variations in the
system parameters, and in general it is more difficult to satisfy the
density criterium than the energy criterium. In practice, we have performed
a large number of BA-LDA calculations within DFT, until finding values 
of $v_i$ that produce approximately the same density profile. In the
vicinity of these BA-LDA values we have then performed a few DMRG calculations, 
for selected values of $U$, to confirm the observed trend. This procedure
is motivated by the large difference in computing time required by both
methods (seconds to minutes, as compared to hours or days, depending on
the system parameters and the required precision).

The data in Fig. \ref{fig2} show the same trend as those in Fig. \ref{fig1},
indicating that the density profile of the superlattice is about one
order of magnitude more sensitive to changes in the potential than in the
interaction, even for quite large values of $U$. This is the same finding
obtained in Ref. \cite{superlatticeprb} for repulsive interactions.
Since the nature of the ground state at attractive and repulsive interactions
is very different, we infer that we are observing a general feature of
periodically modulated systems, not a specific consequence of one type 
of ground state. The ratio between compensating potentials and interactions is,
however, not the same for the energy criterium and for the density criterium, 
and also differs, for each criterium, for attractive and for repulsive 
interactions, which indicates that there is no universal energy scale 
associated with the compensation.

\section{Conclusions}
\label{conclusions}

Density-functional and density-matrix renormalization group calculations
have been performed for superlattices with different modulation patterns,
in the presence of attractive interactions. Results for ground-state
energies and charge-density profiles, obtained
for different lattices and with DFT or DMRG, are mutually consistent and all
point towards the same conclusion: {\em modulations in the on-site potential 
are much more efficient in shaping the system properties than modulations
in the on-site interaction.} This observation has two distinct consequences, 
depending on whether ones interest lies in the design of materials and devices, 
or in the understanding of spontaneously emerging spatial inhomogeneity.

If one wants to design a superlattice with specific electronic properties, e.g.,
as part of a nanoscale device, the key parameter to modulate is the local
electric potential. Spatial variations of the effective particle-particle
interactions are of secondary importance.
If, on the other hand, one is given a system in which nanoscale spatial 
inhomogeneity appears spontaneously, such as in many HTSC cuprates, our results
imply that the analysis of origins and consequences of the modulation in terms
of strong-correlation (large $U$) effects alone is incomplete, and must
necesssarily be accompanied by accounting for local electric fields.

{\bf Acknowledgments}\\
This work was sup\-por\-ted by FAPESP and CNPq.

\end{document}